\documentclass[letterpaper, 10pt, conference, twocolumn]{IEEEtran}
\usepackage{fancyhdr}
\usepackage{epsfig}
\usepackage{threeparttable}
\usepackage{epsf,epsfig}
\usepackage{amsthm}
\usepackage{amsmath}
\usepackage{amssymb}
\usepackage{amsfonts}
\usepackage[noadjust]{cite}
\usepackage{dsfont}
\usepackage{subfigure}
\usepackage{color}

\newtheorem{theorem}{Theorem}

\newtheorem{lemma}{Lemma}
\newtheorem{corollary}{Corollary}

\def\E{\mathsf{E}}

\def\phi{\varphi}

\def\l{\left}
\def\r{\right}
\def\({\left(}
\def\){\right)}

\def\b0{{\mathbf{0}}}

\newcommand{\Pout}{P_{\mathsf{out}}}

\newcommand{\nn}{\nonumber}

\begin{document}
\title{\huge \setlength{\baselineskip}{30pt} A Stochastic-Geometry Approach to Coverage in \\ Cellular Networks with Multi-Cell Cooperation}
\author{
\authorblockN{Kaibin Huang}
\authorblockA{
School of Electr.  \& Electronic Engr.\\
Yonsei University, Korea\\
Email: huangkb@me.com\vspace{-10pt}}
\and
\authorblockN{Jeffrey G. Andrews}
\authorblockA{
School of Electr.  \& Computer Engr.\\
The University of Texas at Austin, USA\\
Email: jandrews@ece.utexas.edu \vspace{-10pt}}} \maketitle

\maketitle
\begin{abstract}
Multi-cell cooperation is  a promising approach for mitigating inter-cell interference in dense cellular networks. Quantifying the performance of  multi-cell cooperation is challenging as it integrates physical-layer techniques and network topologies. For tractability,  existing work typically relies on the over-simplified Wyner-type models. In this paper, we propose a new stochastic-geometry model for a cellular network with multi-cell cooperation, which accounts for practical factors including the irregular  locations of base stations (BSs) and the resultant path-losses. In particular, the proposed network-topology model has three key features: i)  the cells are modeled using a Poisson random tessellation generated by Poisson distributed BSs, ii) multi-antenna BSs are clustered using a hexagonal lattice and BSs in the same cluster mitigate mutual interference by spatial interference avoidance, iii) BSs near cluster edges  access a different sub-channel from that  by other BSs,  shielding  cluster-edge mobiles from strong interference. Using this model and assuming sparse scattering, we analyze the shapes of the outage probabilities of mobiles served by cluster-interior BSs as the average number $K$ of  BSs per cluster increases. The outage probability of a mobile near a cluster center is shown to be proportional to $e^{-c(2-\sqrt{\nu})^2K}$ where $\nu$ is the fraction of BSs lying in the interior of clusters  and $c$ is a constant. Moreover, the outage probability of a typical mobile is proved to scale proportionally with  $e^{-c' (1-\sqrt{\nu})^2K}$ where $c'$ is a constant.
\end{abstract}

\section{Introduction}

Inter-cell interference is the key throughput limiting factor for  dense cellular  networks. A promising approach for mitigating such interference is multi-cell cooperation, namely the joint processing of signals transmitted/received by multiple base stations (BSs) \cite{Gesbert:MultiCellMIMOCooperativeNetworks:2010}. The significant gains promised by multi-cell cooperation have motivated extensive research on the designs of joint transmission techniques (see e.g., 
\cite{DahYu:CoordBeamformMulticell:2010, Huh:MultiuserMISOIntercelInterf:2010}) and the network information-capacity \cite{Hanly:InfoTheoCapacityMultiRxNetworks:1993, Shamai:InfoTheoSymmetricCellularChannel:2002,  Lapidoth:CognitiveNetworkClusteredDecoding:2009, Sanderovich:UplinkMacroDiversityLimitedBackhaul:2009, Simeone:LocalBSCooperationFinite-CapLink:2009}. The scenario of cooperative base stations connected to a central processor via finite-rate backhaul links is considered in \cite{Sanderovich:UplinkMacroDiversityLimitedBackhaul:2009} and the uplink sum rate per cell is derived. Given that neighboring base stations are connected by finite-rate backhaul links, the maximum spatial-multiplexing gain per cell  is derived in \cite{Lapidoth:CognitiveNetworkClusteredDecoding:2009} and observed to be between $0.5$ and $1$. The dependance of this gain on the backhaul-link capacity is characterized in \cite{Simeone:LocalBSCooperationFinite-CapLink:2009}. Most existing work on the network capacity for multi-cell cooperation  is   based on the simple Wyner-type  models, where single-antenna base stations are arranged in a line and interference exists only between neighboring cells \cite{Hanly:InfoTheoCapacityMultiRxNetworks:1993, Shamai:InfoTheoSymmetricCellularChannel:2002}.  The Wyner-type models used in existing work for tractability are oversimplified and fail to account for the random geographical locations of network nodes,  and   the resultant heterogeneous path losses and channel statistics of different links \cite{XuAndrews:AccuracyWynerModel,Gesbert:MultiCellMIMOCooperativeNetworks:2010}.  In particular, the signal-to-interference-and-noise ratio of a mobile is fixed and there is no difference between cell-edge and cell-interior mobiles. In view of prior work, a more practical and versatile network model is needed for addressing some fundamental yet open issues concerning multi-cell cooperation e.g., the effect of the number of cooperative BSs on the network performance and the realistic  performance gains achievable  by BS cooperation. 

Building on \cite{Andrews:TractableApproachCoverageCellular:2010} assuming single-cell transmission,  this paper proposes a stochastic-geometry model for a cellular network with multi-cell cooperation, which accounts for nodes' random  locations and the resultant path losses. BSs  are modeled as a homogeneous Poisson point process (PPP). The cells form a Poisson spatial tessellation generated by  the BS process. Mobiles in each cell are served by the corresponding BS based on time-division multiple access (TDMA). We cluster BSs using a hexagonal lattice and BSs in the same cluster cooperate in transmission. Specifically, each BS employs multi-antennas to avoid interference to single-antenna mobiles served by other cooperative BSs.  To protect cluster-edge mobiles against inter-cluster interference, cluster-edge BSs are assigned a sub-channel  for transmission different from that used by  other BSs.  

In this paper, we focus on the performance of  mobiles served by cluster-interior BSs.  Using the above network model and assuming sparse scattering, the network coverage is quantified by analyzing the outage probabilities of mobiles for a large average number $K$ of cooperative BSs. It is shown that the mobiles near cluster centers have the outage probabilities proportional to $e^{-b\zeta^{\frac{2}{\alpha}}(2-\sqrt{\nu})^2K}$, where $\zeta$ is the response ratio between the main and side lobes of beamformers, $\nu$ the average fraction of BSs in the interior of clusters, $\alpha> 2$ the path-loss exponent, and $b$ a constant. The outage probability of a typical mobile is proved to  decay with increasing $K$ at a slower rate, namely about $e^{-b'\zeta^{\frac{2}{\alpha}}(1-\sqrt{\nu})^2K}$ where $b'$ is a constant. This is consistent with the fact that the cluster-center mobiles are farther away from interferers than a typical mobile. 

{\bf Notation:} The complement of a set $\mathcal{A}$ is represented by $\bar{\mathcal{A}}$. The operator $|\cdot |$  on a vector gives its Euclidean norm and that on a set gives its cardinality. Let $\mathcal{O}(X, r)$ represent a disk centered at $X$ and having the radius $r$. Two functions $f(x)$ and $g(x)$ are asymptotically equivalent if $\frac{f(x)}{g(x)}\rightarrow 1$ as $x\rightarrow \infty$, denoted as $f(x) \sim g(x)$; the cases of $\lim_{x\rightarrow\infty}\frac{f(x)}{g(x)}\geq 1$ and $\lim_{x\rightarrow\infty}\frac{f(x)}{g(x)}\leq 1$ are represented by $f(x) \gtrsim g(x)$ and $f(x) \lesssim g(x)$, respectively. 

\section{Mathematical Models and Metric}
\subsection{Network Topology} 
\begin{figure}
\begin{center}
\includegraphics[width=9cm]{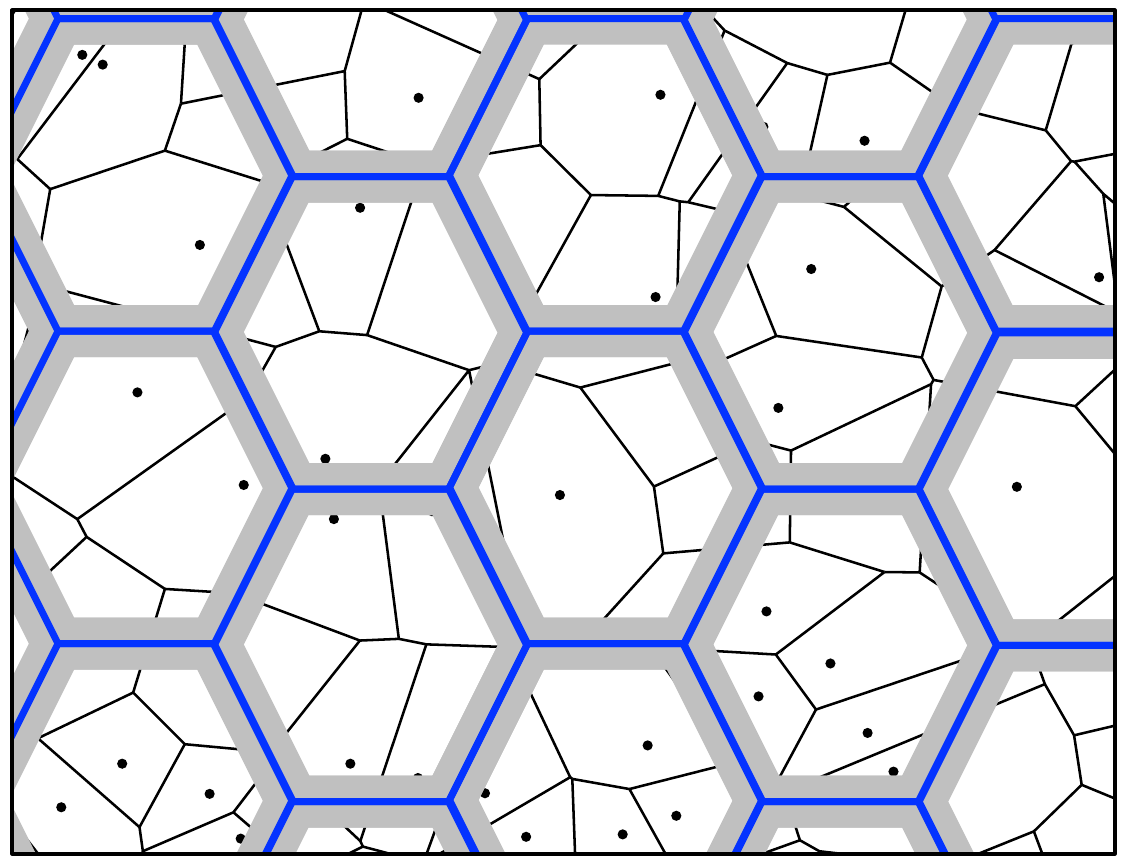}
\caption{The topology of the cellular network with clustered BSs. The cells are drawn using thin black  lines and the cluster regions  thick blue lines; BSs are marked using black dots. The BSs at the cluster edges (shaded area) transmit using a different  sub-channel from  other BSs. }
\label{Fig:NetTopology}
\end{center}
\end{figure}

The BSs are modeled as  a homogeneous PPP $\Phi=\{Y\}$ with the density $\lambda$ where  $Y$ is the coordinates of the represented point.   The  mobiles form a stationary process independent with $\Phi$ and each mobile is assigned to the nearest BS. 
As illustrated in Fig.~\ref{Fig:NetTopology}, the horizontal plane $\mathds{R}^2$ is partitioned into cells using the BS process $\Phi$  and the nearest-neighbor rule. Thus, the cell $\mathcal{V}_Y$ served by the BS $Y\in\Phi$ is   defined as
\begin{equation}\label{Eq:Cell:Def}
\mathcal{V}_Y \!= \l\{A \in \mathds{R}^2\mid |A - Y| \leq |A - X| \ \forall \ X \in \Phi\backslash \{ Y\}\r\}. \!\!\!
\end{equation}
We consider downlink transmission and the  BS $Y$ serves the mobiles in $\mathcal{V}_Y$ based on TDMA. 

\begin{figure}
\begin{center}
\includegraphics[width=9cm]{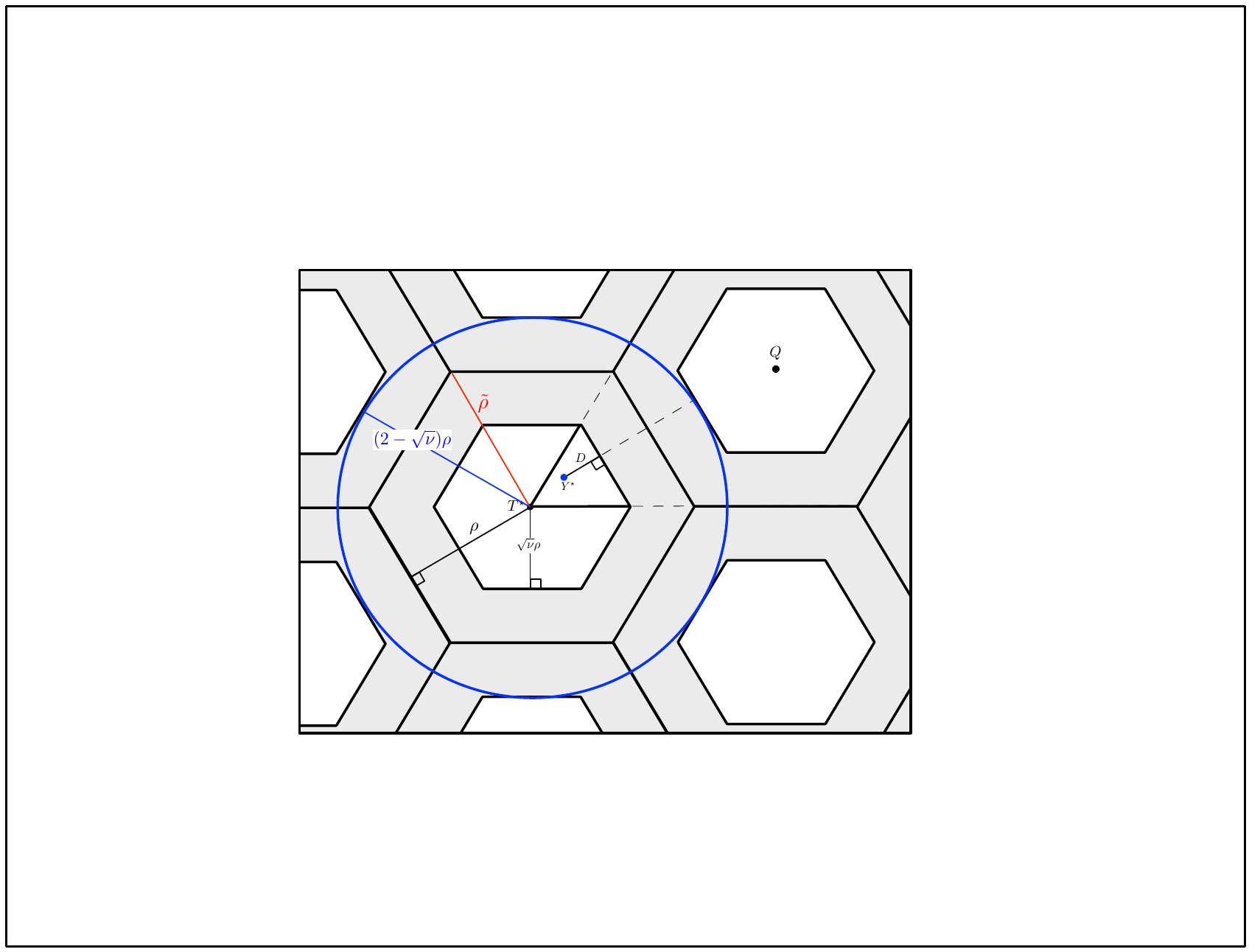}
\caption{The geometric definitions of different  parameters of a typical cluster. In particular, $\rho^2 = \frac{1}{2\sqrt{3}\eta}$ and $\tilde{\rho}^2 = \frac{2}{3\sqrt{3}\eta}$. }
\label{Fig:HexagonBound}
\end{center}
\end{figure}

The BSs are clustered using a hexagonal lattice $\Omega = \{T\}$ with the density $\eta$ and the lattice point $T\in \mathds{R}^2$, modeling  cluster with equal areas. To model non-uniform cluster areas, the lattice can be replaced by a random spatial tessellation, which is currently under investigation. 
Using  the lattice points as the \emph{cluster centers} and applying the nearest-neighbor rule, the horizontal plane is partitioned into hexagonal \emph{cluster regions} as illustrated in Fig.~\ref{Fig:NetTopology}. Let $\mathcal{C}(T, r)$ denote a hexagon centered at $T\in\mathds{R}^2$ and having the distance $r$ from $T$ to an edge. Thus, as illustrated in Fig.~\ref{Fig:HexagonBound},  the cluster region centered at a typical point   $T^\star\in \Omega$ is $\mathcal{C}(T^\star, \rho)$ with  $\rho^2 = \frac{2}{3\sqrt{3}\eta}$ and the area $\lambda/\eta$, enclosing the cluster-interior region $\mathcal{C}(T^\star, \sqrt{\nu}\rho)$ with $0< \nu \leq 1$. The corresponding cluster of cooperative BSs is  $\Phi\cap \mathcal{C}(T^\star, \sqrt{\nu}\rho)$. 

Inspired by the fractional-frequency reuse in WiMax systems, we consider the following frequency reuse scheme. A BS $Y$ in the typical cluster accesses one given  sub-channel if $Y \in \mathcal{C}(T^\star, \rho)\backslash\mathcal{C}(T^\star, \sqrt{\nu}\rho)$ that defines the \emph{cluster edge}, or otherwise transmits in the other sub-channel. Allowing cooperation between appropriately grouped BSs in cluster edges prevents cluster-edge mobiles from receiving strong inter-cluster  interference. Therefore, the analysis in the sequel focuses on the performance of  mobiles served by cluster-interior BSs, which potentially limits the network coverage.

\subsection{Channel Model} We assume narrow-band sub-channels and sparse scattering. Multiple antennas are employed at each BS while mobiles have single antennas and receive single data streams. 
The signals transmitted by the BS $Y$ is received by the mobile $U$ with the power $P_YG_{UY} |U-Y|^{-\alpha}$ where $P_Y$ is transmission power,   $|U-Y|^{-\alpha}$ the path loss,  $G_{UY}$ the spatial response of the beamformer at $Y$ in the direction $(U-Y)$. We consider an interference-dominant  network which  multi-cell cooperation targets, and channel noise is omitted. 

\subsection{Multi-Cell Transmission} 
It is  assumed that symbol boundaries are synchronized. Given TDMA, only a single user is active in each cell and  each slot. We consider multi-cell cooperation using  \emph{interference coordination} that requires potential CSI sharing but no data exchange between cooperative BSs \cite{Gesbert:MultiCellMIMOCooperativeNetworks:2010}. A BS shapes its radiation pattern by spatial filtering such that  a beam is steered towards the intended mobile and no radiation is directed towards  unintended mobiles served by other cooperative BS. Employing $M$ transmit antennas, a BS can   null the interference for up to $(M-1)$ unintended mobiles. Since the numbers of antennas at all BSs are finite, the maximum  of spatial response of the beamformer at an arbitrary BS $Y$, denoted as $W_Y$,  has bounded support $[\delta_1, \delta_2]$ with  $\delta_1, \delta_2 > 0$ and the side-lobes are limited by $\delta > 0$. 


Power control is used at each BS to avert the  path loss. 
Specifically, each BS adapts transmission power such that the receive power is unit.  Let $L_Y$ denote distance from $Y$ to the assigned mobile. Then the transmission power of $Y$ is $P_Y = L_Y^\alpha/W_Y$. Note that the average power is finite since the distribution of $L_Y^\alpha$ has a sub-exponential tail (see \eqref{Eq:L:Dist}) and $W_Y$ is bounded away from zero.

\subsection{Performance Metric} Consider a typical mobile $U^\star$ served by a typical BS  in the typical cluster interior  $\mathcal{C}(T^\star, \sqrt{\nu}\rho)$.  Let $\theta$ denote the required receive signal-to-interference ratio (SIR) for correct decoding. Given power control, the outage probability $\Pout$ for $U^\star$ can be written as $\Pout  = \Pr(I(U^\star) > 1/\theta)$ where $I(U^\star)$ denotes the total interference power received at $U^\star$. 
The outage probability measures the average percentage of mobiles having unsuccessful communication. Alternatively, $(1-\Pout)$ gives the average fraction of mobiles in coverage \cite{Andrews:TractableApproachCoverageCellular:2010}. The direct  analysis of $\Pout$ is intractable due to the complex network topology. Thus, we focus on characterizing the shape of $\Pout$, which is sufficiently accurate for providing useful  design insight. To be specific, $\Pout$ is approximated by an exponential function and the exponent $\psi$ is defined as $\psi = -\log\Pout$ and analyzed in the sequel. 

\section{Outage-Probability Exponent}

\subsection{Main Results} First, consider a typical cluster region $\mathcal{C}^\star$ centered at $T^\star$. The performance of a typical mobile is upper bounded by one near $T^\star$, where  the distances   to the interfering BSs in other clusters are the longest statistically. 

\begin{theorem} \label{Theo:CCMobile:Sparse:PC:FFR} 
\emph{As $K\rightarrow\infty$,  the outage-probability exponent $\psi$ of a mobile located at the center of a typical  cluster region scales as 
\begin{equation}
\lim_{K\rightarrow\infty}\frac{\psi(K)}{\frac{\pi}{2\sqrt{3}}\l(\frac{\delta_1}{\delta\theta}\r)^{\frac{2}{\alpha}}(2-\sqrt{\nu})^2K} = 1. 
\end{equation}}
\end{theorem}

The above result implies that the outage probability of a mobile near the cluster center decays exponentially with $K$ as 
$e^{-c\l(\frac{\delta_1}{\delta}\r)^{\frac{2}{\alpha}}(2-\sqrt{\nu})^2K}$ as $K \rightarrow \infty$, where $c$ is a constant. The scaling also reveals that the outage probability is a sub-exponential function of the ratio $(\delta_1/\delta)$ measuring the sharpness of beams. For the case of no frequency reuse ($\nu =1$), $\Pout \approx e^{-c\l(\frac{\delta_1}{\delta\theta}\r)^{\frac{2}{\alpha}}K}$.

Next, the outage probability of a typical mobile $U^\star$ served by a typical BS $Y^\star$ in the typical cluster interior  $\mathcal{C}(T^\star, \sqrt{\nu}\rho)$ is shown below. 
\begin{theorem} \label{Theo:TypicalMobile:Sparse:PC:FFR} 
\emph{As $K\rightarrow\infty$,  the outage-probability exponent $\psi$ of the  typical mobile  $U^\star$ scales as 
\begin{equation}
\frac{1}{\l[1+4\l(\frac{\delta_1}{\delta\theta}\r)^{\frac{1}{\alpha}}\r]^2}\leq \lim_{K\rightarrow\infty}\frac{\psi(K)}{\frac{2\pi}{\sqrt{3}}\l(\frac{\delta_1}{\delta\theta}\r)^{\frac{2}{\alpha}}(1-\sqrt{\nu})^2 K } \leq  1. 
\end{equation}}
\end{theorem}

\emph{Remark $2.1$}: This result shows that the outage probability of a typical mobile decreases with $K$ exponentially similarly as mobiles  near cluster centers but at a slower rate. Specifically, the factors $(2-\sqrt{\nu})^2$ and $4(1-\sqrt{\nu})^2$ in the two outage-probability exponents  result from the shortest possible distances $(2-\sqrt{\nu})\rho$ and $2(1-\sqrt{\nu})\rho$ between a mobile at a cluster center and a typical mobile with their interferers, respectively. 

\emph{Remark $2.2$}: For the case of no frequency reuse $(\nu = 1)$, Theorem~\ref{Theo:TypicalMobile:Sparse:PC:FFR} reveals that the outage-probability fails to decrease exponentially with $K$. We can show that the outage probability in this case scales with $K$ following a power law. Due to the lack of space, the details will be presented in a separate paper.

\emph{Remark $2.3$}: Noting $K = \lambda/\eta$, the results in Theorem~\ref{Theo:CCMobile:Sparse:PC:FFR} and \ref{Theo:TypicalMobile:Sparse:PC:FFR} show the advantage of employing more BSs in the cellular network provided that the average number of cooperative BSs  increases accordingly. This is in contrast with the case of single-cell transmission where increasing the BS density gives no performance gain as observed in \cite{Andrews:TractableApproachCoverageCellular:2010}.

\subsection{Proofs}
\subsubsection{Proof of Theorem~\ref{Theo:CCMobile:Sparse:PC:FFR}}
Before proving the theorem, it is useful to characterize the truncated shot noise. Specifically, define the truncated shot noise  $\hat{I}(X, r)$ measured at the location $X$ as a shot noise truncated by the disk $\mathcal{O}(X, r)$:
\begin{equation}\label{Eq:TrunIPower}
\hat{I}(X, r) = \sum_{Y \in \Phi\cap \bar{\mathcal{O}}(X, r)} P_YG_{XY} |X-Y|^{-\alpha}.
\end{equation}
The interference power of a mobile  given Poisson distributed interferers is also studied in \cite{Ganesh:LargeDeviationInterferenceWirelessNet} in the setting of a mobile ad hoc network, where the tail-probability exponents of the interference power are derived using large deviation theory. The following corollary can be proved using a similar procedure as \cite[Theorem~$12.$iv)]{Ganesh:LargeDeviationInterferenceWirelessNet}. 

\begin{corollary}\label{Cor:Ganesh} \emph{Let $G$ denote a random variable following the common distribution of  $\{P_YG_{XY}\}$. If $-\log\Pr(G> x) \sim cy^\gamma$ with $c > 0$ and $0< \gamma < 1$, the tail-probability exponent of  $\hat{I}(X, r)$ scales as follows:
\begin{equation}
1\leq \lim_{r\rightarrow\infty} \frac{-\log\Pr(\hat{I}(X, r) > x)}{cr^{\alpha \gamma}x^\gamma } \leq 2^\alpha.\nn
\end{equation}}
\end{corollary}

\begin{lemma}\label{Lem:TXPwr:Tail} The distribution of $P_YG_{XY}$ has a sub-exponential tail: \[-\log \Pr( P_YG_{XY} > x) \sim \pi\lambda\l(\delta_1/\delta\r)^{\frac{2}{\alpha}} x^{\frac{2}{\alpha}}.\] 
\end{lemma}
This matches the condition stated in Corollary~\ref{Cor:Ganesh}. The proof is provided in the appendix. 

\noindent{\bf Proof of Theorem~\ref{Theo:CCMobile:Sparse:PC:FFR}.} Let $\mathcal{B}$ denote the union of all cluster-interior regions containing the interferers for $U^\star$ served by a BS $Y^\star\in \mathcal{C}(T^\star, \sqrt{\nu}\rho)$. As observed from  Fig.~\ref{Fig:HexagonBound},  $\mathcal{B}$ can be bounded as 
\begin{equation}
\mathcal{H} \subset \mathcal{B}  \subset \bar{\mathcal{O}}(T^\star, (2 - \sqrt{\nu})\rho) \label{Eq:Cluster:Bnds}
\end{equation}
where $\mathcal{H} = \mathcal{C}(T^\star, (2-\sqrt{\nu})\rho+\epsilon)\cap\mathcal{C}(Q, \sqrt{\nu}\rho)$ and $\mathcal{C}(Q, \sqrt{\nu}\rho)$ is the cluster-interior region next to the typical one. 
Using the above result, the interference power for $U^\star$ conditioned on $U^\star = T^\star$ is bounded as 
\begin{equation}
\Lambda\leq I(T^\star) \leq \hat{I}(T^\star, (2 - \sqrt{\nu})\rho) \nn
\end{equation}
where \[\Lambda=[(2 - \sqrt{\nu})\rho+\epsilon]^{-\alpha}\sum_{Y\in\Phi\cap\mathcal{H} }P_YG_{T^\star Y}.\] 
It follows that 
\begin{eqnarray}
\Pr(\hat{I}(T^\star, (2\! -\! \sqrt{\nu})\rho) > \theta^{-1}) &\geq& \Pout(U^\star = T^\star)\label{Eq:Proof:c4:UB}\\
&\geq&\Pr(\Lambda \theta > 1). \label{Eq:Proof:c4:LB}
\end{eqnarray}
Using Corollary~\ref{Cor:Ganesh} and Lemma~\ref{Lem:TXPwr:Tail}, the exponent of the upper bound in \eqref{Eq:Proof:c4:UB} scales as 
\begin{equation}
\lim_{\rho\rightarrow\infty}\frac{-\log\Pr(\hat{I}(T^\star, (2 - \sqrt{\nu})\rho) > \theta^{-1})}{\pi\lambda \l(\frac{\delta_1}{\delta\theta}\r)^{\frac{2}{\alpha}}\rho^2} \geq 1. \nn
\end{equation}
Given $\rho^2 = \frac{1}{2\sqrt{3}\eta}$ and $K = \lambda/\eta$, 
\begin{equation}
\lim_{K\rightarrow\infty}\frac{-\log\Pr(\hat{I}(T^\star, (2 - \sqrt{\nu})\rho) > \theta^{-1})}{\pi\l(\frac{\delta_1}{\delta\theta}\r)^{\frac{2}{\alpha}}(2 - \sqrt{\nu})^2K} \geq \frac{1}{2\sqrt{3}}. \label{Eq:Proof:c5}
\end{equation}
By applying Lemma~\ref{Lem:TXPwr:Tail} and \cite[Lemma~$2.2$]{AsmussenBook:RuinProb} and  letting $\epsilon \rightarrow 0$, the lower bound in \eqref{Eq:Proof:c4:LB} can be shown to scale as 
\begin{equation}
\lim_{K\rightarrow\infty}\frac{-\log \Pr(\Lambda> \theta^{-1})}{\pi\l(\frac{\delta_1}{\delta\theta}\r)^{\frac{2}{\alpha}}(2 - \sqrt{\nu})K} = \frac{1}{2\sqrt{3}}. \label{Eq:Proof:c6}
\end{equation}
Combining \eqref{Eq:Proof:c4:UB}, \eqref{Eq:Proof:c4:LB},  \eqref{Eq:Proof:c5}, and \eqref{Eq:Proof:c6} gives the desired result. \hfill $\blacksquare$

\subsubsection{Proof of Theorem~\ref{Theo:TypicalMobile:Sparse:PC:FFR}} The claim  in the theorem statement is proved by combining the following two lemmas and substituting $\rho^2 = \frac{1}{2\sqrt{3}\eta}$ and $K = \lambda/\eta$. 
\begin{lemma}\label{Lem:TypicalMobile:Sparse:PC:FFR:LB} \emph{As $\rho\rightarrow\infty$,  the outage-probability exponent $\psi$ of a typical  mobile satisfies 
\begin{equation}
\psi(\rho) \gtrsim \frac{\l(\frac{\delta_1}{\delta\theta}\r)^{\frac{2}{\alpha}}4\pi\lambda (1-\sqrt{\nu})^2\rho^2}{\l[1+2\l(\frac{\delta_1}{\delta\theta}\r)^{\frac{1}{\alpha}}\r]^2}. 
\end{equation}}
\end{lemma}
\begin{lemma}\label{Lem:TypicalMobile:Sparse:PC:FFR:UB} \emph{As $\rho\rightarrow\infty$,  the outage-probability exponent $\psi$ of a typical mobile  satisfies 
\begin{eqnarray}
\psi(\rho) \lesssim \l(\frac{\delta_1}{\delta\theta}\r)^{\frac{2}{\alpha}}4\pi\lambda (1-\sqrt{\nu})^2\rho^2 .
\end{eqnarray} }
\end{lemma}

The proofs of  the above lemmas are presented in the appendix. 

\bibliographystyle{ieeetr}

\appendix 
\noindent{\bf Proof of  Lemma~\ref{Lem:TXPwr:Tail}}  For convenience, define $\beta = W_Y/G_{XY}$ with the support $[\delta_1/\delta, \infty)$ and the distribution function $f_\beta$. Since $P_Y = L_Y^\alpha/W_Y$ and \cite{Andrews:TractableApproachCoverageCellular:2010}
\begin{equation}\label{Eq:L:Dist}
\Pr(L_Y \geq \tau) = e^{-\pi \lambda \tau^2}, 
\end{equation}
we can upper bound $\Pr( P_YG_{XY} > x)$ as follows:
\begin{eqnarray}
&&\hspace{-27pt}\Pr( P_YG_{XY} > x) \nn\\
&=& \E\l[e^{-\pi\lambda (\beta x)^{\frac{2}{\alpha}}}\r]\nn\\
&=& \int^\infty_{(1+\epsilon)\frac{\delta_1}{\delta}} e^{-\pi\lambda \l(\tau x\r)^{\frac{2}{\alpha}}}f_\beta(\tau)d\tau + \nn\\
&& \int^{(1+\epsilon)\frac{\delta_1}{\delta}}_{\frac{\delta_1}{\delta}} e^{-\pi\lambda \l(\tau x\r)^{\frac{2}{\alpha}}}f_\beta(\tau)d\tau, \quad \epsilon > 0\label{Eq:Proof:c1}\\
&\leq& e^{-\pi\lambda \l((1+\epsilon)\frac{\delta_1 x}{\delta} \r)^{\frac{2}{\alpha}}}\Pr\l(\beta > {(1+\epsilon)\frac{\delta_1}{\delta}}\r) + \nn\\
&&e^{-\pi\lambda \l(\frac{\delta_1 x}{\delta} \r)^{\frac{2}{\alpha}}}\Pr\l(\frac{\delta_1}{\delta}\leq \beta \leq 
(1+\epsilon)\frac{\delta_1}{\delta}\r). \nn
\end{eqnarray}
It follows that 
\begin{equation}
\lim_{x\rightarrow\infty}\frac{-\log \Pr( P_YG_{XY} > x)}{\pi\lambda \l(\frac{\delta_1 }{\delta} \r)^{\frac{2}{\alpha}}x ^{\frac{2}{\alpha}}} \leq 1. \label{Eq:Proof:c2}
\end{equation}
From \eqref{Eq:Proof:c1}, 
\begin{eqnarray}
\Pr( P_YG_{XY} > x) 
&\geq& \int^{(1+\epsilon)\frac{\delta_1}{\delta}}_{\frac{\delta_1}{\delta}} e^{-\pi\lambda \l(\tau x\r)^{\frac{2}{\alpha}}}f_\beta(\tau)d\tau\nn\\
&\geq& e^{-\pi\lambda \l((1+\epsilon)\frac{\delta_1 x}{\delta} \r)^{\frac{2}{\alpha}}}\times\nn\\
&&\Pr\l(\frac{\delta_1}{\delta}\leq \beta \leq 
(1+\epsilon)\frac{\delta_1}{\delta}\r). \nn
\end{eqnarray}
Thus, 
\begin{equation}
\lim_{x\rightarrow\infty}\frac{-\log \Pr( P_YG_{XY} > x)}{\pi\lambda \l(\frac{\delta_1 }{\delta} \r)^{\frac{2}{\alpha}}x ^{\frac{2}{\alpha}}} \geq (1+\epsilon)^{\frac{2}{\alpha}}.  \label{Eq:Proof:c3}
\end{equation}
Combining  \eqref{Eq:Proof:c2} and \eqref{Eq:Proof:c3} and letting  $\epsilon \rightarrow 0$ gives the desired result. 
\hfill $\blacksquare$

\noindent {\bf Proof of Lemma~\ref{Lem:TypicalMobile:Sparse:PC:FFR:LB}.}
Consider a typical mobile $U^\star$ served by a BS $Y^\star$ in the typical cluster-core region $\mathcal{C}(T^\star, \sqrt{\nu}\rho)$. 
Due to frequency reuse, the BSs interfering with  $U^\star$ lie outside the hexagon  $\mathcal{C}(T^\star, (2-\sqrt{\nu})\rho)$ (see Fig.~\ref{Fig:HexagonBound}).   Thus the interference power measured at $U^\star$ can be upper bounded as
\begin{eqnarray}
\hspace{-10pt}I(U^\star) &\leq& \sum_{Y \in \acute{\Phi}} P_YG_{U^\star Y} |Y - U^\star|^{-\alpha} \nn\\
&\leq& \sum_{Y \in \acute{\Phi}} P_YG_{U^\star Y} \l[\max\l(|Y - U^\star|, L \r)\r]^{-\alpha}\label{Eq:Proof:a7}\\
&\leq& \sum_{Y \in \acute{\Phi}} P_YG_{U^\star Y} \l[\max\l(|Y - Y^\star|-L, L \r)\r]^{-\alpha}\label{Eq:Proof:a1}
\end{eqnarray}
where  $\acute{\Phi} = \Phi \cap \bar{\mathcal{C}}(T^\star, (2-\sqrt{\nu})\rho)$, 
\eqref{Eq:Proof:a7} follows from \eqref{Eq:Cell:Def} that prevents any  interferer to be nearer to $U^\star$ than the serving BS, and  \eqref{Eq:Proof:a1} applies the triangular inequality. 

Let $\mathcal{C}^o$  and $\mathcal{C}^i$ denote the boundary and interior of $\mathcal{C}(T^\star, (2-\sqrt{\nu})\rho)$, respectively, where $\mathcal{C}^o = \bar{\mathcal{C}}^i \cap\mathcal{C}(T^\star, (2-\sqrt{\nu})\rho)$. Define the distance $D$ of a typical  BS $Y^\star$ in $\mathcal{C}(T^\star, \sqrt{\nu}\rho)$ to its boundary as $D = \min_{X \in \mathcal{C}^o} |X - Y^\star|$ (see Fig.~\ref{Fig:HexagonBound}). Using the property that $Y^\star$ is uniformly distributed in $\mathcal{C}(T^\star, \sqrt{\nu}\rho)$, the distribution of $D$ can be obtained as 
\begin{equation}\label{Eq:D:Dist}
\Pr(D \geq d) = \l(1 - \frac{d}{\sqrt{\nu}\rho}\r)^2, \quad 0 \leq d \leq \sqrt{\nu}\rho. 
\end{equation}
From the definition of $D$, the disk 
$\mathcal{O}(Y^\star, D')$ with $D' = D+2(1-\sqrt{\nu})\rho$ belongs to $\bar{\mathcal{C}}(T^\star, (2-\sqrt{\nu})\rho)$ (see Fig.~\ref{Fig:HexagonBound}). It follows from this fact and \eqref{Eq:Proof:a1} that 
\begin{eqnarray}
I(U^\star) &\leq& \sum_{Y \in \mathcal{D}} P_YG_{U^\star Y} \l[\max\l(|Y - Y^\star|-L, L \r)\r]^{-\alpha}\nn\\
&=& \sum_{Y \in \mathcal{D}} P_YG_{U^\star Y} \l(|Y - Y^\star|-L\r)^{-\alpha}, \quad D' > 2L \nn\\
&=& J(Y^\star, D', L). \label{Eq:Proof:a2}
\end{eqnarray}
where $\mathcal{D} = \Phi \cap \bar{\mathcal{O}}(Y^\star, D')$. 
It can be observed that 
\begin{equation}
\begin{aligned}
&\Pr(J(Y^\star, D', L)\geq x \mid D' > 2L) \\ 
=&\Pr(\hat{I}(Y^\star, D' - L)\geq x \mid D' > 2L).  \label{Eq:Proof:a3}
\end{aligned}
\end{equation}
It follows from  \eqref{Eq:Proof:a2} and  \eqref{Eq:Proof:a3} that 
\begin{equation}
\begin{aligned}
&\Pr(I(U^\star) > x\mid D' > 2L ) \\
\leq&\Pr(\hat{I}(Y^\star, D' - L)\geq x \mid D' > 2L). \label{Eq:Proof:a5}
\end{aligned}
\end{equation}
For $0 < \tau < 1$, expanding $\Pr(I(U^\star) > x)$ gives 
\begin{eqnarray}
&&\hspace{-25pt}\Pr(I(U^\star) > x) \\
&\leq&\Pr(I(U^\star) > x\mid D + 2(1-\sqrt{\nu})\tau\rho> 2L ) +\nn\\
&&\Pr(D + 2(1-\sqrt{\nu})\tau\rho\leq 2L)\nn\\
&\leq& \Pr(\hat{I}(Y^\star, D' - L)\geq x \mid D + 2(1-\sqrt{\nu})\tau\rho> 2L) + \nn\\
&&\Pr(D + 2(1-\sqrt{\nu})\tau\rho\leq 2L)\label{Eq:Proof:a5a}\\
&\leq& \Pr(\hat{I}(Y^\star, 2(1-\sqrt{\nu})(1-\tau)\rho)\geq x) +\nn \\
&&\Pr(D + 2(1-\sqrt{\nu})\tau\rho\leq 2L)\nn
\end{eqnarray}
where \eqref{Eq:Proof:a5a} uses \eqref{Eq:Proof:a5} and the fact that the event $\{D + 2(1-\sqrt{\nu})\tau\rho> 2L\}$ implies $\{D' \geq 2L\}$.  Using \eqref{Eq:L:Dist} and given $\epsilon > 0$, 
\begin{eqnarray}
&&\hspace{-25pt}\Pr(D + 2(1-\sqrt{\nu})\tau\rho\leq 2L)\nn\\
 &=& \E\l[e^{-\frac{\pi\lambda}{4}\l(D + 2(1-\sqrt{\nu})\tau\rho\r)^2}\r]\nn\\
&=& \int_{\epsilon\sqrt{\nu}\rho}^{\sqrt{\nu}\rho} e^{-\frac{\pi\lambda}{4}\l(x + 2(1-\sqrt{\nu})\tau\rho\r)^2} f_D(x) dx + \nn\\
&& e^{-\frac{\pi\lambda}{4}\l(2(1-\sqrt{\nu})\tau\rho\r)^2} \Pr( D\leq \epsilon\sqrt{\nu}\rho)\nn
\end{eqnarray}
where the last equation uses the distribution of $D$  in \eqref{Eq:D:Dist}.  It follows that 
\begin{eqnarray}
-\log\Pr(D + 2(1-\sqrt{\nu})\tau\rho\leq 2L)\sim \frac{\pi\lambda}{4}\l(2(1-\sqrt{\nu})\tau\rho\r)^2. \label{Eq:Proof:a10}
\end{eqnarray}
From Corollary~\ref{Cor:Ganesh} and Lemma~\ref{Lem:TXPwr:Tail}, 
\begin{equation}
\begin{aligned}
&\ -\log \Pr(\hat{I}(Y^\star, 2(1-\sqrt{\nu})(1-\tau)\rho)\geq \theta^{-1}) \\
\gtrsim&\ \pi\lambda \l(\frac{\delta_1}{\delta\theta} \r)^{\frac{2}{\alpha}} \l(2(1-\sqrt{\nu})(1-\tau)\rho\r)^2. 
\end{aligned}\label{Eq:Proof:a11}
\end{equation}
Combining \eqref{Eq:Proof:a5a}, \eqref{Eq:Proof:a10} and \eqref{Eq:Proof:a11} and applying \cite[Lemma~$1.2.15$]{DemboBoo:LargeDeviation} gives 
\begin{equation}
\begin{aligned}
&\ -\log \Pr(I(U^\star) > \theta^{-1})\\
 \gtrsim&\ \min\l(\tau^2,  4 \l(\frac{\delta_1}{\delta\theta}  \r)^{\frac{2}{\alpha}}(1-\tau)^2\r)\pi\lambda(1-\sqrt{\nu})^2\rho^2. 
\end{aligned}\nn
\end{equation}
Maximizing the above lower bound on $-\log \Pr(I(U^\star) > x)$ over $\tau$ gives the desired result. 
\hfill $\blacksquare$

\noindent {\bf Proof of Lemma~\ref{Lem:TypicalMobile:Sparse:PC:FFR:UB}.}  Since $Y^\star$ is uniformly distributed in $\mathcal{C}(T^\star, \sqrt{\nu}\rho)$, $Y^\star$ can be constrained to lie in the triangle $\Delta\in\mathcal{C}(T^\star, \sqrt{\nu}\rho)$ defined below without affecting the distribution of $D$: (see Fig.~\ref{Fig:HexagonBound})
\begin{equation}
\Delta = \{X\in \mathcal{C}(T^\star, \sqrt{\nu}\rho) \mid 0 < \angle(X - T^\star)\leq 60^o\}. 
\end{equation}
Let $Q$ denote the point in the cluster-center lattice that is one of the nearest to $T^\star$, namely $|Q-T^\star| = 2\rho$, and  $\angle(Q - T^\star) = 30^o$ (see Fig.~\ref{Fig:HexagonBound}). Since $\Phi\cap \mathcal{C}(Q, \sqrt{\nu}\rho)$ gives a subset of interferers for $U^\star$,  $I(U^\star)$ can be lower bounded as 
\begin{eqnarray}
I(U^\star) &\geq& \sum_{Y \in \Phi \cap \mathcal{C}(Q, \sqrt{\nu}\rho)} P_YG_{U^\star Y} \l(|Y - Y^\star|+L\r)^{-\alpha}\nn\\
&\geq& [2(1-\sqrt{\nu})\rho + D + L ]^{-\alpha} \hspace{-10pt}\sum_{Y \in \Phi \cap \mathcal{C}(Q, \sqrt{\nu}\rho)} P_YG_{U^\star Y}.  \nn
\end{eqnarray}
Thus, 
\begin{eqnarray}
&&\hspace{-25pt}\Pr(I(U^\star) > \theta^{-1})\nn\\
&\geq& \Pr\l(\sum_{Y \in \mathcal{F}} P_YG_{U^\star Y}  > \theta^{-1}(2(1-\sqrt{\nu})\rho + D + L )^{\alpha}\r)\nn\\
&\geq& \Pr\l(\sum_{Y \in \mathcal{F}} P_YG_{U^\star Y}  > \theta^{-1}(2(1-\sqrt{\nu})\rho + D + L )^{\alpha}\mid \r.\nn\\
&&\l.L = \ell, D = d\r) \Pr(L \leq \ell)\Pr(D \leq d)\label{Eq:Proof:a8}
\end{eqnarray}
where $\mathcal{F} = \Phi \cap\mathcal{C}(Q, \sqrt{\nu}\rho)$. 
Combining   \eqref{Eq:Proof:a8} with \cite[Lemma~$2.2$]{AsmussenBook:RuinProb} and \eqref{Eq:D:Dist} gives the desired result. 
  \hfill$\blacksquare$

\end{document}